\documentclass{nature}
\usepackage{epsfig}
\usepackage{amsmath}
\usepackage{amssymb}
\usepackage{color}

\begin{document}

\title{Ultra-high mechanical stretchability and controllable topological phase transitions in two-dimensional arsenic}

\author{Ming Yang and Wu-Ming Liu$^{\star}$}
\maketitle

\begin{affiliations}
\item
Beijing National Laboratory for Condensed Matter Physics,
Institute of Physics, Chinese Academy of Sciences,
Beijing 100190, China

$^\star$e-mail: wliu@iphy.ac.cn

\end{affiliations}

\begin{abstract}
  The mechanical stretchability is the magnitude of strain which a material can suffer before it breaks.
  Materials with high mechanical stretchability, which can reversibly withstand extreme mechanical deformation and cover arbitrary surfaces and movable parts, are used for stretchable display devices, broadband photonic tuning and aberration-free optical imaging. 
  Strain can be utilised to control the band structures of materials and can even be utilised to induce a topological phase transition, driving the normal insulators to topological non-trivial materials with non-zero Chern number or Z2 number.
  Here, we propose a new two-dimensional topological material with ultra-high mechanical stretchability - the ditch-like 2D arsenic. This new anisotropic material possesses a large Poisson's ratio $\nu_{xy}=1.049$, which is larger than any other reported inorganic materials and has a ultra-high stretchability 44\% along the armchair direction, which is unprecedent in inorganic materials as far as we know. Its minimum bend radius of this material can be as low as 0.66 nm, which is comparable to the radius of carbon-nanotube.
  Such mechanical properties make this new material be a stretchable semiconductor which could be used to construct flexible display devices and stretchable sensors.
  Axial strain will make a conspicuous affect on the band structure of the system, and a proper strain along the zigzag direction will
  drive the 2D arsenic into the topological insulator in which the topological edge state can host dissipation-less spin current and spin transfer toque, which are useful in spintronics devices such as dissipation transistor, interconnect channels and spin valve devices.
\end{abstract}


The mechanical stretchability is the magnitude of strain which a material can suffer before it fails, which is one of few most important parameters for solid materials.
For those material with large flexibility, strain tuning becomes a more effective method to tailor the material's electronic properties. Materials with high mechanical stretchability can reversibly withstand extreme mechanical deformation and cover arbitrary surfaces and movable parts. They are used in the situations such as stretchable display devices, aberration-free optical imaging, and epidermal sensing.
When strain is applied, the Poisson's ratio relates the resulting lateral strain to applied axial strain, and is another important mechanical parameters in designing micro devices\cite{Poisson_nanowires,Negative_Poisson}. The applications of Poisson¡¯s ratio include casing design and wellbore stability analysis. For typical inorganic materials, the Poisson's ratio is smaller than 0.5, for example, silicon: 0.22, copper: 0.355. So it is interesting to search for the those system in which the flexibility and the Poisson's ratio is much larger.
The applying and the adjusting of the strains will modify the atomic coordinates and induce the changes of the hopping parameters, which will alter the electronic band-structure and even induce a topological phase transition.

The topological property of the material is a topic of great interest due to the robust backscattering-free transport and the unique spin-texture pattern in its surface state.
Among all the means (such as impurity doping) which can intrigue topological phase transition, strain tuning have the advantage of can be continuously turned, conferring people more  degree of freedom to control the topological property of the materials.
New materials in which topological phase transitions could be controlled by strain, are considered to be valuable and receive a good amount of attentions.
In this letter, we propose a new material, single layer arsenic for the first time,
show that it has a very large Poisson's ratio and an ultra-high mechanical flexibility along the armchair direction, and demonstrate that topological property of this material could be tailored  by external strain and this material could become topological insulator.

The bulk orthorhombic As has an obviously layered structure\cite{arsenic_bulk}, as is shown in Fig. 1 (a).
The new 2D material we proposed is shown in Fig. 1 (b). It is just a single layer exfoliated from the bulk structure, as is denoted by the blue box in Fig. 1 (a). Such exfoliation method has successfully used to obtain graphene\cite{Graphene_exfoliation1} and phosphorene\cite{Phosphorene_exfoliation}.
The top-view and side-view of the primitive cell are shown in Fig. 1 (c) and (d).
The in-plane bond-angle and bond-length are denoted by $\alpha$ and $l_1$ respectively, while the inter-plane bond-angle and bond-length are represented by $\beta$ and $l_2$ respectively.
By variable-cell relaxations, we obtain the optimized lattice parameters of this ditch-like 2D As are a = 3.690 {\AA}, b = 4.765 {\AA}, which are close to the lattice parameters of the bulk As. Considering that this new 2D Arsenic has a ditch-like puckered hexagonal structure and is similar to the phosphorene, we call it as arsenene in the following of this paper.

Then, we calculated the phonon dispersion
 for the ditch-like arsenene by the finite displacement method.
Fig. 1 (f) shows the phonon dispersion
for arsenene at its natural state.
The phonon dispersion shows no imaginary frequency, indicating
that arsenene is stable.
The dynamical stability of the material is further
checked by finite temperature molecular dynamics simulations at
temperature 300 K for room temperature and 30 K for low temperature.
During the simulations, a 4x4 super-cell containing 64
atoms is used. The length of time-step is chosen as 5 fs and simulations
with 1000 steps are executed. It is observed that, the atoms
shake around the equilibrium positions back and forth while the
extent of such motion under 300 K is larger than under 30 K ( the
evolution of atomic positions can be found in movies in Extended Data
). However, no structural collapse happens throughout
the simulations, which can also be seen from the free energies
curves as the functions of time-step shown in Fig. 1 (g). It is also
observed that, the crystal structure always remains nearly the same
as the initial crystal structure. Actually, as is shown in the inset of
Fig. 1 (g), the crystal structure corresponding to the last free energy
maximum in T=300 K case (right), still shows no significant structural
differences as compared with the initial crystal structure (left).
The lattice relaxation, phonon modes
analysis together with FTMD simulations mentioned above provide
an authentic test for the stability of arsenene.

We further analysed 2D As with several other structures with higher symmetry, including graphene like planar structure with 6-fold rotation symmetry, triangle structure with 3-fold rotation symmetry and square structure with 4-fold rotation symmetry. The phonon dispersions of these competing structures show negative frequencies and hence couldn't stably exist (see Extended Figure 1). Consequently, in the following we focus on the buckled hexagonal structures which could stably exist.

We impose the uni-axial strain on the system, to calculate the Poisson's ratios. After imposing the strain to one axis, the lattice constant of the other axis was fully relaxed through the technique of energy minimization to ensure the force in the transverse direction is a minimum. (a) illustrates how the length of a-axis change when imposing strain on b-axis. (b) gives the changes of the length of b-axis with the imposing of strain on a-axis. The $\delta a=a/a_{0}-1$ and $\delta b=b/b_{0}-1$ represent the change ratio of a- and b-axis.
A linear property can be clearly seen. In fact, the linearly dependent coefficients are higher than 98\%, while the red lines are the linear fit of the data.
From the slopes of the fit line we can arrive at the Poisson's ratio as:
$\nu_{yx}=0.3336$, $\nu_{xy}=1.0492$.
The Poisson's ratio $\nu_{xy} (1.0492)$ found here, as far as we know, is larger than any other inorganic substance. For comparison, the Poisson's ratio of several common material is listed as follows: concrete (0.20), steel (0.27), copper (0.33), gold (0.44), graphene (0.18).
We verify the above-mentioned Poisson's ratio of our system by the Quantum Espresso code and find the difference of the Poisson's ratio calculated by VASP and Quantum Espresso are less than 1\%, and such accordance proofs that our system indeed possesses large Poisson's ratio.

With the purpose of investigating the stretchability of arsenene, we calculate the stress as a function of the strain, as is shown in Fig. 2 (c)-(d). In Fig. 2 (c)-(d), the black curves corresponds to the undoped case, red curves represent the 0.05 hole/cell doped case, green curves represent the 0.05 electron/cell doped case. Fig. 2 (c) and (d) shows respectively the effect of the uni-axial strain along the zigzag direction and along the armchair direction. From the place where stress maximum/minimun occur we can find the stretchability of the material, such method is successfully used in predicting the stretchability of graphene. It can be seen from Fig. 2 (c) that, the stretchability along a- (zigzag) axis is 13\% for press, and 21\% for stretch. Doping 0.05 hole per unit-cell will make the press stretchability increase to 15\% and doping 0.05 electron per unit-cell will increase the stretch strain to 22\%. It can be seen from Fig. (d) that, the stretchability along b- (armchair) axis is 22\% for press, and 44\% for stretch. Such large stretchability, as far as we know, have not appeared in other inorganic materials.

In order to understand why such material can suffer from such a large strain without dissociation, we analysed how the bond-length and the bond-angle change when applied an axial strain. The definition of the in-plane (inter-plane) bond-angle and bond-length are given in Fig. 1 (c) and (d).
It can be seen that, under the same strain, the bond-length shows an much less apparent change than the bond-angle. Take the b-axis (armchair) strain as an example,when imposing 10\% tensile strain, the inter-layer bond-angle changes only 4.97\%, while the inter-layer bond-length changes 0.418\%, much more than the relative change of the bond-length. That's to say, the change of bond-angle greatly relieves the change of bond-length and makes the material being stable even under large magnitude of strain.


We now analyse the influence of the band structures of arsenene and illustrate how to acquire the topological insulating phase. Based on the relaxed structures, we calculated the band structures of arsenene at its natural state and after imposing stains. Fig.  4 (a)-(d) depict the band structures of arsenene at 0\%, 6\%,12\% and 12.1\% uniaxial strain along zigzag direction, respectively. It is clear that, before imposing strain, arsenene is indirect-gap insulator (Fig. 4 (a)), with its conduction band minimum (CBM) located at $\Gamma$ point and valence band maximum (VBM) located on Y-$\Gamma$ path. With the increase of the strain along the zigzag direction, the band structure gradually become direct-gap type, with both the CBM and VBM located at $\Gamma$ point.  As an example, we plotted the the band structure at 6\% strain along the zigzag direction in Fig. 4 (b). When further increase the strength of the uni-axial strain, the size of gap become smaller and smaller. The CBM and VBM will touch with each other at strain 12\%, as is shown in Fig. 4 (c). The system behaves as half metal with point fermi surface.
Fig. 4 (d) shows the band structure under 12.1\% strain, it can be seen that further increase of the strain above 12\% will cause the gap re-opening, which appeared in topological phase transition in $Bi_2Se_3$ and silicene.

To explore the orbital property of the arsenene which have undergone the above-mentioned gap re-opening process, we give the orbital-projected band structures in Fig. 4 (e)-(h), correspond to s, $p_z$, $p_x$ ,$p_y$ orbitals of As element respectively. It can be clearly seen that, at those k-points away from $\Gamma$ point, the top-most valence band is dominated by $p_x$ orbital and the bottom-most conduction band is dominated by $p_z$ orbital. However, at the vicinity of $\Gamma$ point, the orbital component of the these two bands exchange with each other. This evident band inversion is the indication of topological phase transition. To verify the topological property of the system after the band inversion process, we calculated the Z2 index according to the parities of each occupied band by the Fu-Kane criteria\cite{Fu}. The index for two-dimensional topological insulators $v$ is expressed as $(-1)^{v_{0}}=\prod_{i=1}^{4}\delta_{i}$
 in which $\delta_{i}=\prod_{m=1}^{N}\xi_{2m}(\Gamma_{i})$ represents the product of the parities of each occupied band at 4 time-reversal invariant momenta $\Gamma_{i}$.
Extended Data Table 1 lists the parities of each band at every time-reversal-invariant-momentum (TRIM). The '+' and '-' represent positive and negative parity respectively. The products of the occupied bands at each time-reversal invariant momentum are listed in the right-most column. As is shown, the product of parities of occupied bands contributes a $+1$ at ($\pi$,$\pi$) while $-1$ at the three other time-reversal invariant momenta, resulting in $Z2=1$, corresponding to topological insulating phase. This shows the topological phase transition could be controlled by the external strain.

The inter-layer hopping parameter is also an important quantity that will make an influence on the electronic band-structures\cite{HSTao}. By tuning the distance between two layer of arsenic, the inter-layer hopping parameter could be adjusted. In this paragraph, we demonstrate that, by the inclusion of the inter-layer hopping, the strain needed to trigger the topological transition can be reduced to 4\%. Extended Data Figure 2 gives the band-structures under 4\% zigzag compression, while the layer distances are not the same. Extended Figure 2 (a) shows the band-structure of the large layer-distance case, in which the layer-distance is 9.85 {\AA}. In this case the coupling between two layer is so small that the double-layer system behaves as two standalone layer of material. When diminishing the layer-distance to 4.27 {\AA}, the size of energy gap decrease to 0.27 eV (see Extended Figure 2 (b)). When the layer-distance is reduced to 3.71 {\AA}, the conduction band minimum touches the valence band maximum and the gap vanishes. If we further reduce the layer-distance to 3.28 {\AA}, the band inversion occurs and the system becomes a topological insulator, with its band structures shown in Extended Data Fig.2 (d).

In what follows, we discuss how much bending can the arsenene bear, and also investigate the effect of the bending on the electronic band structures.
The ability of a material to withstand the binding can be characterized by the maximal curvature $\kappa=1/r_{min}$ , where $r_{min}$ is the minimal radius of the material before if breaks.
The bent arsenene with a certain curvature can be seen as a cylinder (see Fig. 4 (a) for illustration), and
the building block of the such nanotube structure is the unit-cell of arsenene.
Here, we use the index $N$ to denote the number of unit-cell of arsenene which is contained by the unit-cell of the nanotube.
Then, the number of atoms in a unit-cell of nanotube structure with index N is $4N$, considering the unit-cell of arsenene contains 4 atoms (shown in Fig. 1 (c)-(d)).
Fig. 4 (b)-(f) gives the crystal structures and phonon spectrums of the bent arsenene with different curvatures.
From the phonon spectrums we can see that, the arsenene is stable in case that curvature is less than 1.52 nm$^{-1}$ (corresponding to $N\geq8$).
When curvature is larger than 1.73 nm$^{-1}$ (corresponding to $N\leq7$), the spectrum shows negative frequency near the $\Gamma$ point (indicated by the green circles) and the structures are no longer stable.
The maximal curvature corresponds to the radius 6.6$\AA$, which is comparable to that of the carbon nanotube\cite{c_nanotube1}.
Such small radius indicates the flexibility of the arsenene is very large and this semiconductor could be used as the detector which is able to cover arbitrary surfaces.
Fig. 4 (g) illustrates the free energy belong to each atom as a function of the curvature (or equivalently the index $N$).
It is depicted that, the greater the curvature is, the higher the energy becomes.
This means the arsenene will maintain its 2D structure and do not tend to bend in free-standing state.
Fig. 4 (h) shows the electronic band-structures of the bent arsenene, taking $N=8$ structure as an example.
The electronic energy gap is  0.69 eV, which is similar with the bulk gap of the un-bent arsenene.


Now we give experimental protocols. Considering the structure of the ditch-like arsenene is similar to the bulk structure of As, the arsenene could be obtained by the electrochemical exfoliation\cite{Graphene_exfoliation1} or mechanical exfoliation\cite{Graphene_exfoliation2,Phosphorene_exfoliation} method.
The Poisson's ratio of the material could be electromechanically measured by the lateral atomic force microscope method\cite{Poisson_nanowires}. The topological property of the material could be examined by the transport method.
Similar to Bi$_2$Se$_3$, the observation of the spin-Hall current\cite{Dora} and the non-equally spaced Landau levels\cite{QKXue1} in arsenene will be signatures of the Dirac fermions in surface of the topological insulator\cite{BHZ}.

In conclusion, we have proposed a new 2D material arsenene and verified its stability. Calculation shows that, this material possesses a large Poisson's ratio and ultra-high mechanical stretchability. This material can be very flexible and the minimal radius can be nanometer level. It is found that, External strain could intrigure topological phase transition and change this material to be topological insulator. Our work provide a realistic material with superb mechanical properties and controllable topological phase transition which will have potential applications in spintronics devices such as dissipation transistor, interconnect channels and spin valve devices.

\section*{Methods}

Our first principle calculations are in the framework of the generalized gradient
approximation (GGA) of the density functional theory.
Convergence tests with respect to energy cutoff and k points sampling are performed to ensure numerical accuracy of total energy. We find an energy cutoff of 400 eV and Gamma centered Monkhorst-Pack grids of 10$\times$10$\times$1 for k point sampling are enough to converge the difference in total energy less than 1 meV. The relaxations are carefully made so that the forces on atoms are smaller than 0.001 eV/{\AA} , in which the conjugate gradient algorithm is utilised.
In the finite temperature molecular dynamics simulations, a 4x4 supercell is used and the length of time-step is chosen as 5 fs. The phonon dispersion curve and phonon density of states are obtained using the force-constant method by phonopy code\cite{phonopy}. After the structural relaxation, the SOC is included self-consistently within the second variational method. The DFT codes used in the calculations are Quantum Espresso\cite{QE} and VASP\cite{VASP2}.



\begin{addendum}

\item [Acknowledgement]
We acknowledge helpful discussions with X. X. Wu and X. L. Zhang
This work is supported by the NKBRSFC (Grants Nos. 2011CB921502, 2012CB821305),
NSFC (grants Nos. 61227902, 61378017, 11434015),
SKLQOQOD under grants No. KF201403, SPRPCAS under grants No. XDB01020300.
The numerical calculations are performed on the Shenteng supercomputer at CNIC-CAS and on the TianHe-1A supercomputer at NSCC-tj.

\item [Author Contributions]
M. Y. performed the numerical calculations.
All authors analyzed the data and wrote the manuscript.

\item [Competing Interests]
The authors declare that they have no competing financial interests.

\item [Correspondence]
Correspondence and requests for materials should be addressed to Wu-Ming Liu.
\end{addendum}

\clearpage

\newpage
\bigskip
\textbf{Figure 1 Crystal structures, Brillouin zone and the structural stability of single layer Arsenic.}
(a) The structure of bulk orthorhombic As. (b) The structure of arsenene (single layer puckered hexagonal As). It can be seen that, bulk orthorhombic As has an obviously layered structure, and arsenene is precisely one layer of the bulk As [as is shown by the blue box in (a)]. This means one could obtain arsenene from bulk orthorhombic As by the exfoliating method.
(c) and (d) are the top-view and side-view of the primitive cell, where
the in-plane bond-angle and bond-length are denoted by $\alpha$ and $l_1$ respectively, while the inter-plane bond-angle and bond-length are represented by $\beta$ and $l_2$ respectively.
(e) Brillouin zone. (f) Phonon dispersion with an optical-acoustic separation is observed. The phonon dispersion shows no imaginary frequency which indicates that the system is statical stable. (g) Finite temperature molecular dynamical simulations (FTMD). The red and blue curves represent the free energy with the evolution of time. The curve of 300K oscillates more violently than that of 30K. Although the curves oscillate around their average value, there is no sharp jump shown on the curve. This corresponds to the atoms vibrating around their equilibrium position back and forth with no structural phase transition happened. Actually, the final structure after the simulation process [shown in the right inset of figure 1(f)] shows no major changes as is compared with the initial structure (shown in the left inset of figure 1(f)). Such FTMD results also demonstrate that the single layer Arsenic is stable.

\textbf{Figure 2 Poisson's ratio and stress-strain relation.}
(a) illustrates how the length of a-axis change when imposing strain on b-axis. (b) gives the changes of the length of b-axis with the imposing of strain on a-axis. The $\delta a=a/a_{0}-1$ and $\delta b=b/b_{0}-1$ represent the change ratio of a- and b-axis. The black dots is the data derived by the VASP code, while the red lines are the linear fit of the data.
From the slopes of the fit line we can arrive at the Poisson's ratio as:
$\nu_{yx}=0.3336$, $\nu_{xy}=1.0492$.
The Poisson's ratio $\nu_{xy} (1.0492)$ found here, as far as we know, is larger than any other inorganic substance.
(c) (d) are the stress-strain relations. The black, red and green curves corresponds the undoped case, 0.05 hole per unit-cell doped case and  0.05 electron per unit-cell doped case, respectively. (c) stress-strain relation along zigzag direction.  From the place where stress maximum/minimun occur, we can read out the compressive stretchability along zigzag direction is 12\% for undoped case and doping with 0.05 hole per unit-cell will increase this value to 15\%. We can also read out the tensile stretchability along the zigzag direction is 21\% and this value could be increased to 22\% by doping 0.05 electron per unit-cell.
(d) stress-strain relation along arm-chair direction. We can read out the compressive stretchability along arm-chair direction is 22\% for undoped case. The tensile stretchability along the zigzag direction is 44\%, and such large flexibility, as far as we know, have not appeared in other inorganic materials.
(e) shows how bond-angles will change  when a-axis (zigzag) strain is applied.
(f) displays the changes of the bond-lengths as functions of a-axis (zigzag) strain.
(g) demonstrates how bond-angles will change  when b-axis (armchair) strain is applied.
(h) illustrates the dependence of the bond-lengths on the b-axis (armchair) strain.
$\alpha$ is the intra-layer bond-angle while $\beta$ is the inter-layer bond-angle.
bond-length 1 is the intra-layer bond-length while bond-length 2 is the inter-layer bond-length.
The range of y-axis in all the subfigures are chosen as -15\% to +15\% for the convenience of comparing the magnitude of the relative changes of each structural parameters.
It can be seen that, under the same strain, the bond-length shows an much less apparent change than the bond-angle. Take the b-axis (armchair) strain as an example,when imposing 10\% tensile strain, the inter-layer bond-angle changes only 4.97\%, while the inter-layer bond-length changes 0.418\%, much more than the relative change of the bond-length. That's to say, the change of bond-angle greatly relieves the change of bond-length and makes the material being stable even under large magnitude of strain.



\textbf{Figure 3 Band structures of arsenene under different strain.}
(a) the band structure without strain. An indirect gap is observed with the conduction band minimum located at $\Gamma$ point and the valence band maximum appeared at Y-$\Gamma$  path. (b) the band structure with 6\% strain along the zigzag direction. The band structure become direct-gap type, with both the CBM and VBM located at $\Gamma$ point. (c) The band structure under 12\% strain along the zigzag direction, the gap vanish and the system becomes a zero-gap insulator with conduction band and the valence band degenerate at $\Gamma$ point. (d) The band structures under 12.1\% strain. Gap re-opens when strain excess 12\%. The inset shows the detail of band structure at the vicinity of the Fermi level and $\Gamma$ point.
(e)-(h) are the orbital projected band structures under 12.1\% strain. Purple, blue, red, green represent the s, $p_z$, $p_x$ ,$p_y$ orbitals of As atoms respectively. The radii of circles are proportional to the weight of the corresponding orbitals. It's clear that, in the energy range [-0.5,0.5] eV range, s and $p_y$ orbitals contribute much less than $p_z$ and $p_x$ orbitals. At those k-points far away from $\Gamma$ point, the band contributed by $p_z$ orbital lies above the Fermi level E$_F$. However, this band goes across the E$_F$ and lies below E$_F$ at $\Gamma$ point, showing an $p_z-p_x$ type band inversion at $\Gamma$ point.
(i) s,p,d resolved density of states, where red, green and blue correspond to s,p and d states. It could be seen that, the state near the fermi level is mainly contributed by p-orbitals.
(j) the $p_x$, $p_y$, $p_z$ resolved density of state, where red, green and blue correspond to $p_x$, $p_y$, $p_z$ states respectively.

\textbf{Figure 4 The bending properties of arsenene.}
(a) illustrates the structure of the bent arsenene and how arsenene can fold to form a nanotube structure.
Here, we use the index $N$ to denote the number of unit-cell of arsenene which is contained by the unit-cell of the nanotube.
(b)-(f) gives the crystal structures and phonon spectrums of the bent arsenene with different curvatures.
From the phonon spectrums we can see that, the arsenene is stable in case that curvature is less than 1.52 nm$^{-1}$ (corresponding to $N\geq8$).
When curvature is larger than 1.73 nm$^{-1}$ (corresponding to $N\leq7$), the spectrum shows negative frequency near the $\Gamma$ point (indicated by the green circles) and the structures are no longer stable.
Such large curvature indicates the flexibility of the arsenene is very large and this semiconductor could be used as the detector which is able to cover arbitrary surfaces.
(g) illustrates the free energy belong to each atom as a function of the curvature (or equivalently the index $N$).
It is depicted that, the greater the curvature is, the higher the energy becomes.
This means the arsenene will maintain its 2D structure and do not tend to bend in free-standing state.
Fig. (h) shows the electronic band-structures of the bent arsenene corresponding to $N=8$ structure.
The electronic energy gap is  0.69 eV, which is similar with the bulk gap of the un-bent arsenene.

\textbf{Extended Data Figure 1 Instability of several high symmetry competing structures.}
Crystal structure and the corresponding phonon spectrum of graphene-like planar structure with 6-fold rotation symmetry ((a)-(b)), triangle structure with 3-fold rotation symmetry ((c)-(d)) and square structure with 4-fold rotation symmetry ((e)-(f)).
The phonon dispersions of these competing structures show negative frequencies and hence couldn't stably exist. As a result,
in the manuscript we focus on the ditch-like hexagonal structures which could stably exist.

\textbf{Extended Data Figure 2 Layer-distance-dependent band-structures.}
(a)-(d) are the crystal structures and the corresponding band-structures of double layer arsenic with layer-distance 9.85 {\AA}, 4.27 {\AA}, 3.71 {\AA} and 3.28 {\AA}, at 4\% zigzag compression. The '+' and '-' denote the positive and the negative parity of the band at $\Gamma$ point.  (a) shows the band-structure of the large layer-distance case, in which the layer-distance is 9.85 {\AA}. In this case the coupling between two layer is so small that the double-layer system behaves as two standalone layer of material. (b) When diminishing the layer-distance to 4.27 {\AA}, the size of energy gap decrease to 0.27 eV. (c) When the layer-distance is reduced to 3.71 {\AA}, the conduction band minimum touches the valence band maximum and the gap vanishes. (d) If we further reduce the layer-distance to 3.28 {\AA}, the topological band inversion occurs and the system becomes a topological insulator (Z2=1).



\newpage
\begin{figure}
\begin{center}
\epsfig{file=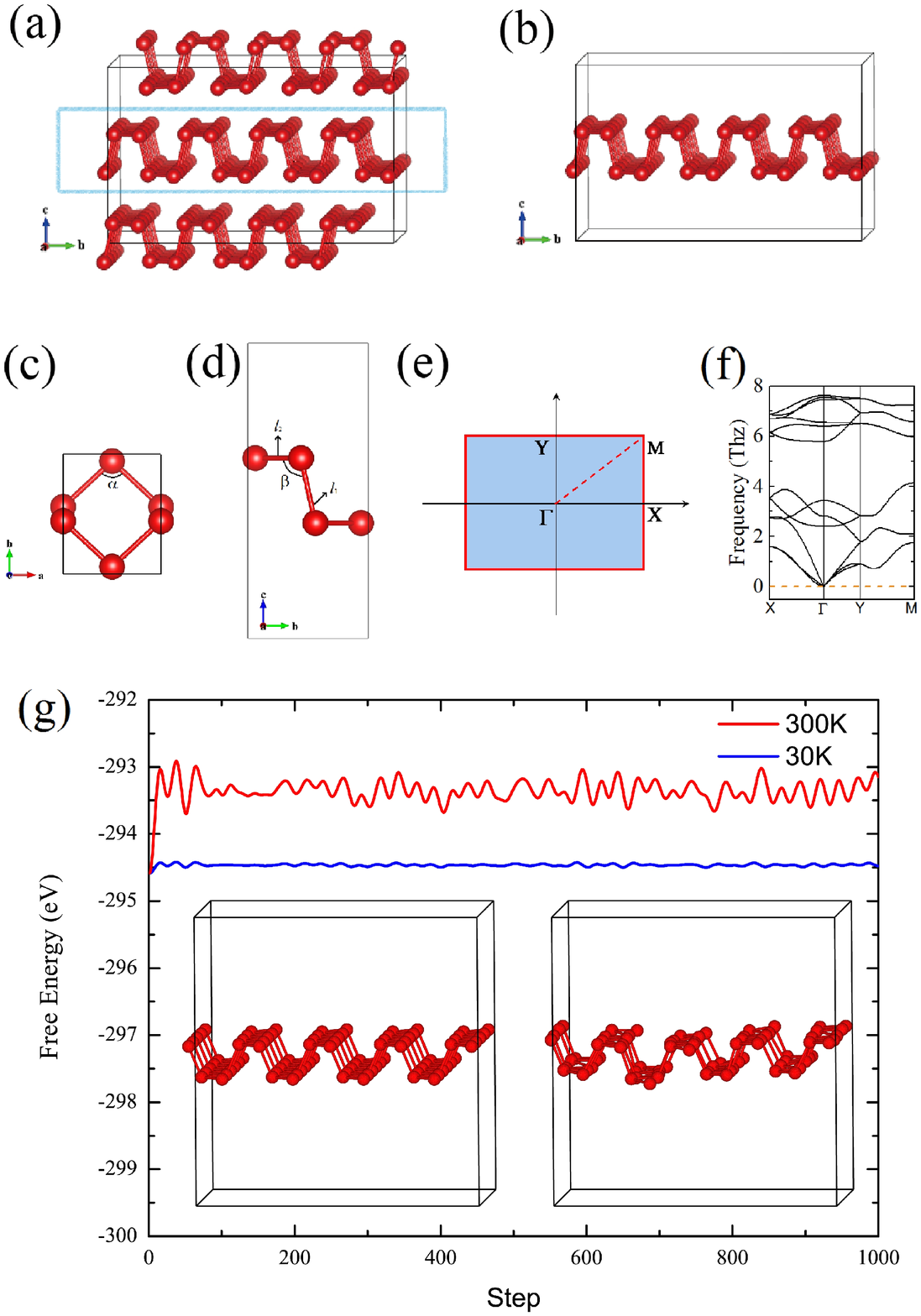,width=15cm}
\end{center}
\label{fig:structure}
\end{figure}

\newpage
\begin{figure}
\begin{center}
\epsfig{file=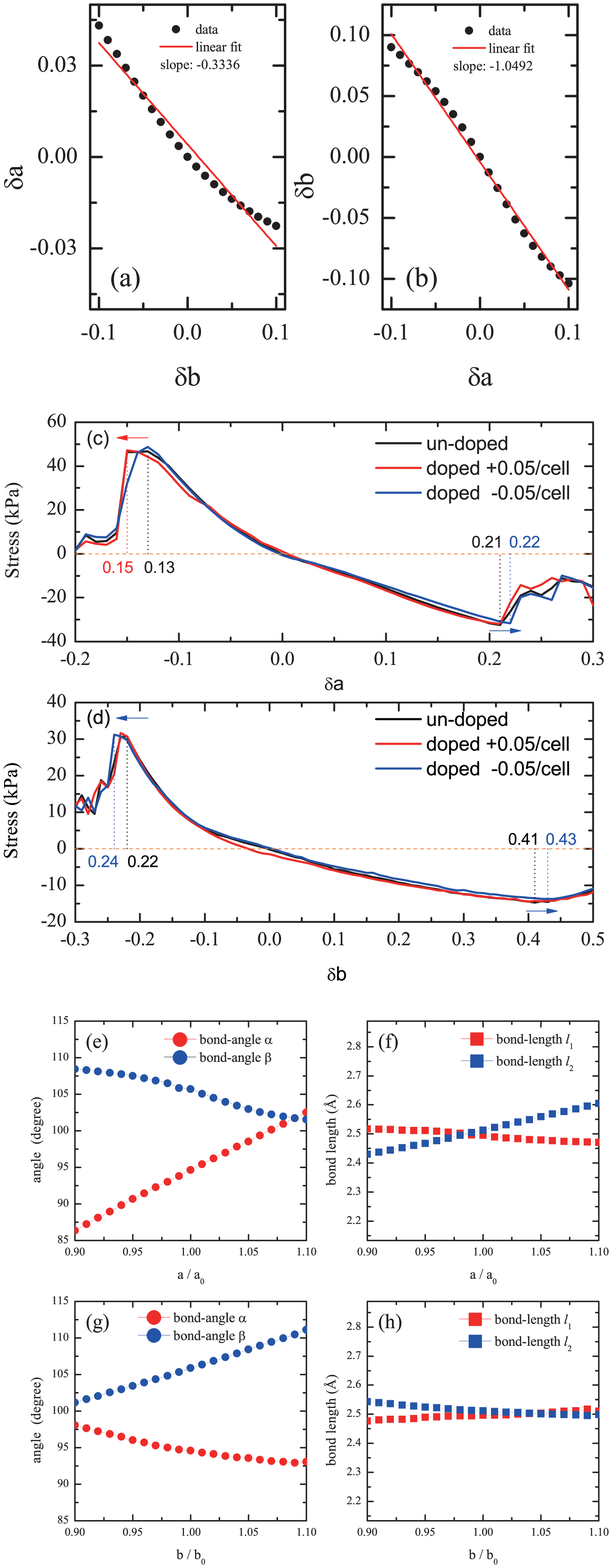,width=8cm}
\end{center}
\label{fig:structure}
\end{figure}

\newpage
\begin{figure}
\begin{center}
\epsfig{file=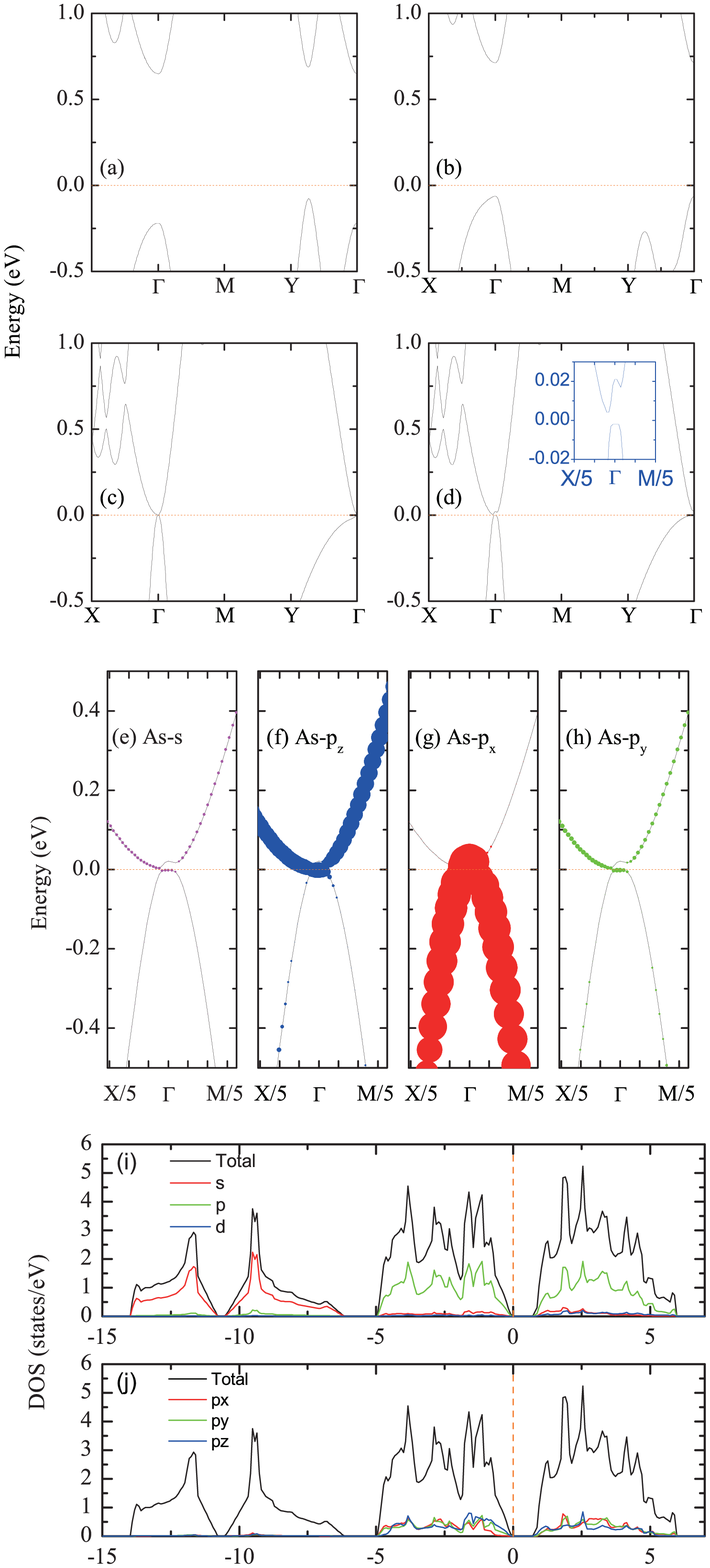,width=10cm}
\end{center}
\label{fig:structure}
\end{figure}

\newpage
\begin{figure}
\begin{center}
\epsfig{file=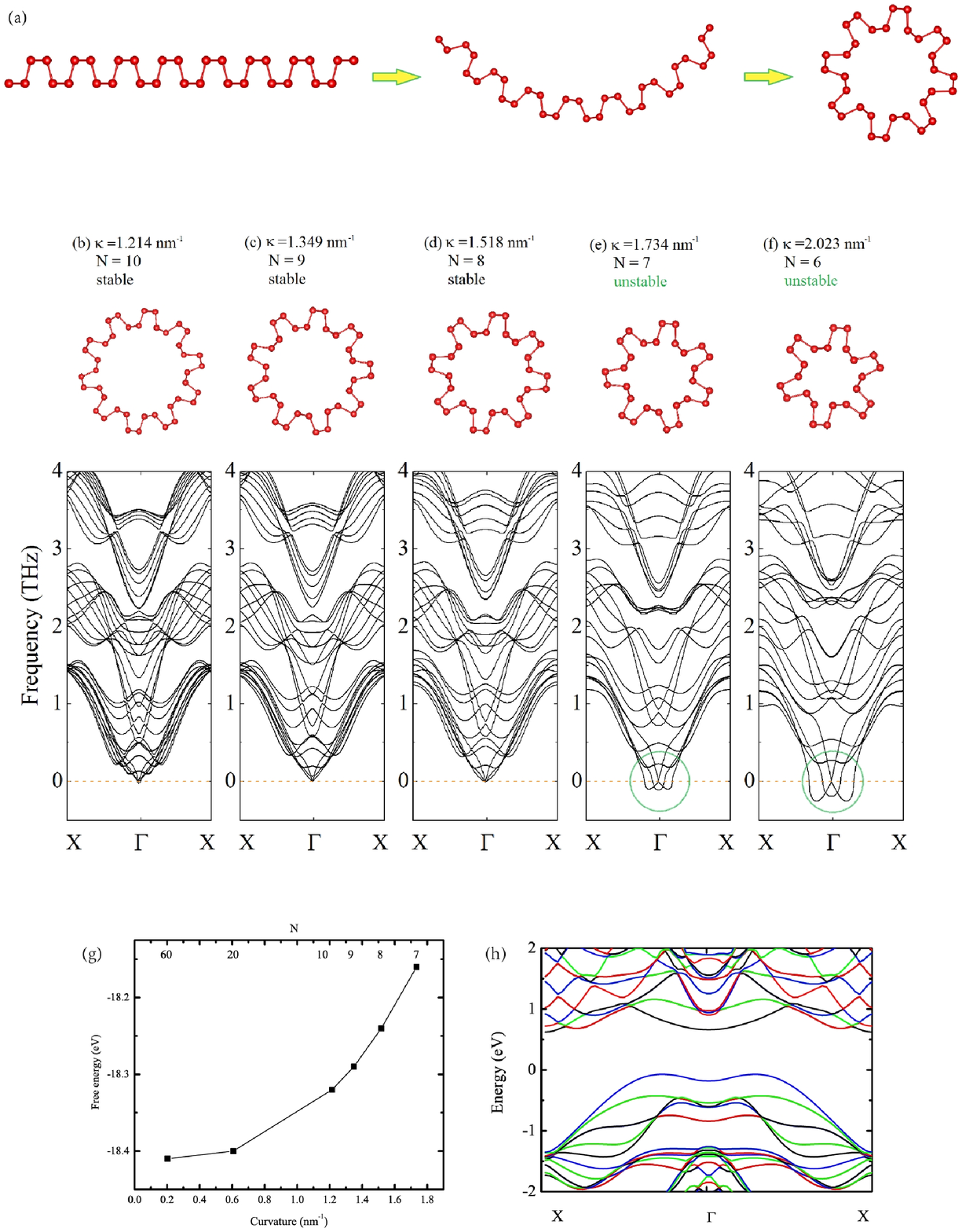,width=16cm}
\end{center}
\label{fig:structure}
\end{figure}

\newpage
\begin{figure}
\begin{center}
\epsfig{file=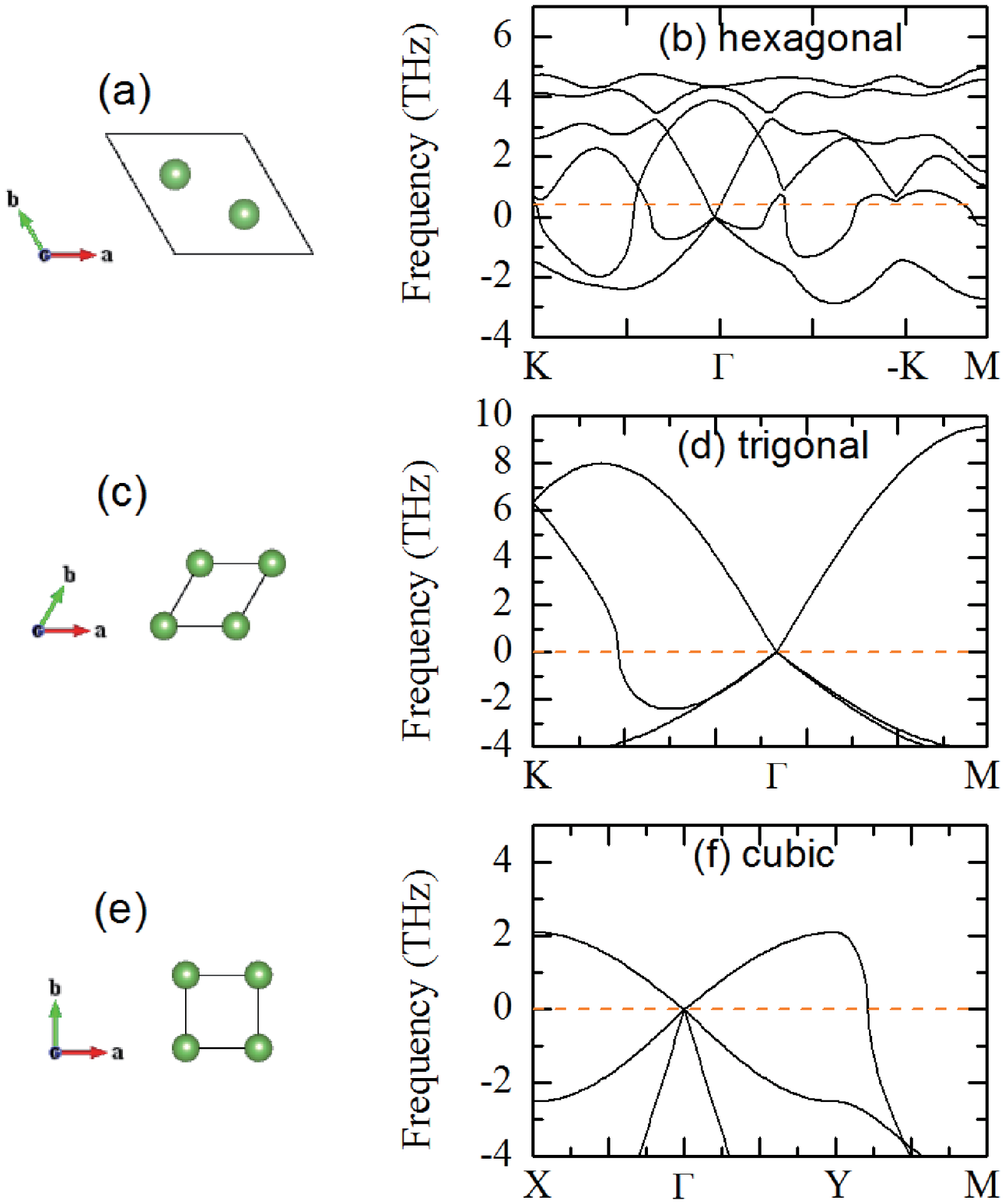,width=15cm}
\end{center}
\label{fig:structure}
\end{figure}

\newpage
\begin{figure}
\begin{center}
\epsfig{file=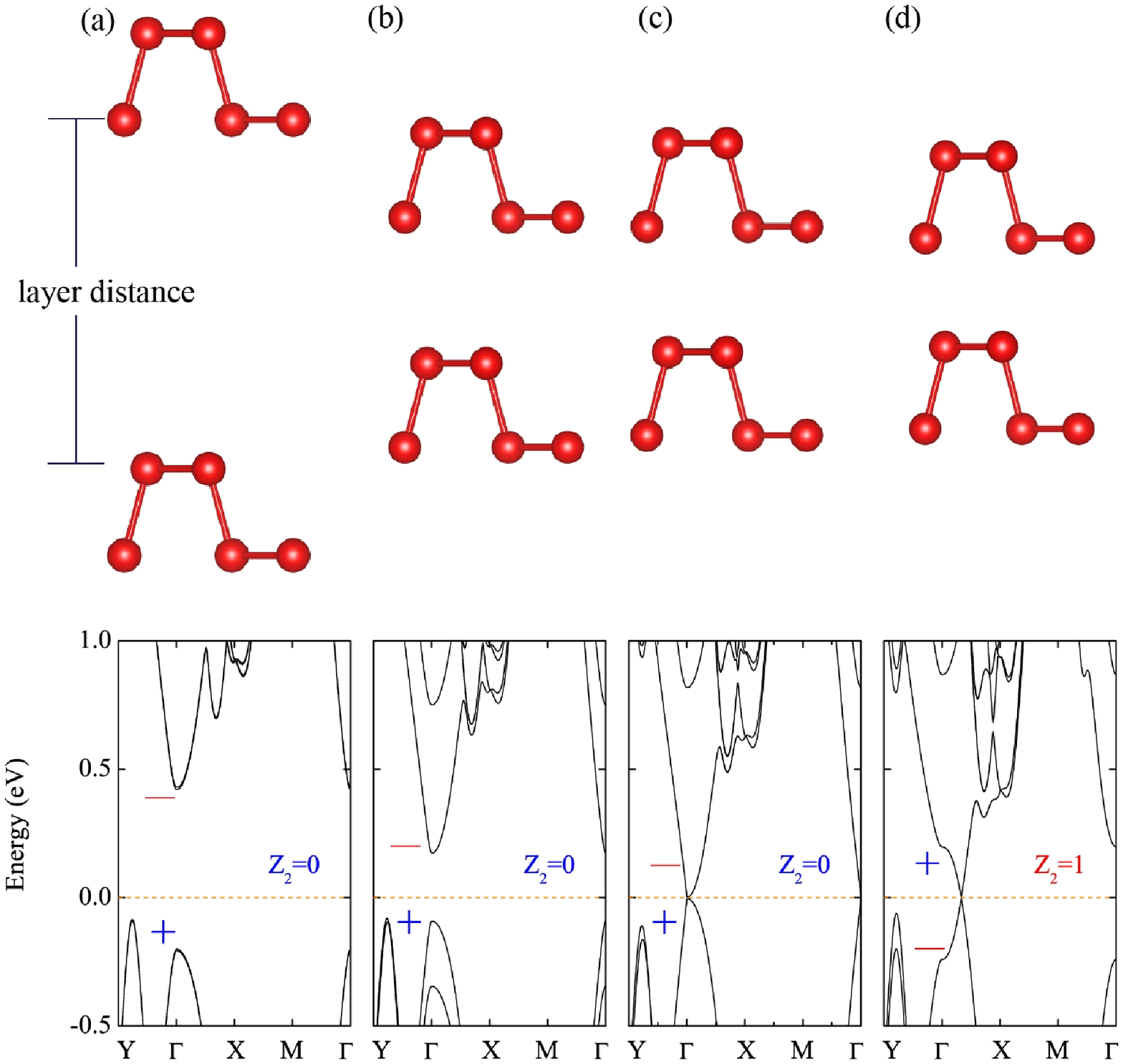,width=12cm}
\end{center}
\label{fig:structure}
\end{figure}

\newpage
\begin{table*}[ht]
\caption{Parities of top-most isolated valence bands at four time-reversal invariant momenta (TRIM). Positive parity is denoted by $+$ while negative denoted by $-$. Products of the parities of the occupied bands at each time-reversal invariant momentum are listed in the right-most column. As is shown, the product of parities of occupied bands contributes a $+1$ at ($\pi$,$\pi$) while $-1$ at the three other time-reversal invariant momenta, resulting in $Z2=1$, corresponding to topological insulating phase. } 
\begin{tabular}[t]{c|c|c}
\hline
TRIM&Parity&Product of the parities\\
\hline
(0,0)&${\color{blue}+}{\color{red}-}{\color{blue}+}{\color{red}-}{\color{blue}+}{\color{red}-}{\color{blue}+}{\color{red}-}{\color{blue}+}{\color{red}-}$&\textbf{${\color{red}-}$}\\
($\pi$,0)&${\color{blue}+}{\color{red}-}{\color{red}-}{\color{blue}+}{\color{blue}+}{\color{red}-}{\color{red}-}{\color{blue}+}{\color{blue}+}{\color{red}-}$&\textbf{${\color{red}-}$}\\
(0,$\pi$)&${\color{blue}+}{\color{red}-}{\color{blue}+}{\color{red}-}{\color{blue}+}{\color{red}-}{\color{red}-}{\color{blue}+}{\color{blue}+}{\color{red}-}$&\textbf{${\color{red}-}$}\\
($\pi$,$\pi$)&${\color{blue}+}{\color{blue}+}{\color{red}-}{\color{red}-}{\color{red}-}{\color{red}-}{\color{blue}+}{\color{blue}+}{\color{blue}+}{\color{blue}+}$&\textbf{${\color{blue}+}$}\\

\hline
\end{tabular}
\label{table:parities} 
\end{table*}


\begin{thebibliography}{99}

\bibitem{BHZ} Bernevig, B. A., Hughes, T. A. \& Zhang, S. C.
Quantum spin Hall effect and topological phase transition in HgTe quantum wells.
\textit{Science} \textbf{314}, 1757 (2006).

\bibitem{LinHsin} Lin, H., Wray, L. A., Xia, Y., Xu, S. Y., Jia, S., Cava, R. J., Bansil, A. \&  Hasan, M. Z.
Half-Heusler ternary compounds as new multifunctional experimental platforms for topological quantum phenomena.
\textit{Nat. Mater.} \textbf{9}, 546 (2010).



\bibitem{Fang_Bi2Se3} Zhang, H. J., Liu, C. X., Qi, X. L., Dai, X., Fang Z. \& Zhang, S. C.
Topological insulators in Bi2Se3, Bi2Te3 and Sb2Te3 with a single Dirac cone on the surface.
\textit{Nat. Phys.} \textbf{5}, 438 (2009).

\bibitem{Fu} Fu, L. \& Kane, C. L.
Topological insulators with inversion symmetry.
\textit{Phys. Rev. B} \textbf{76}, 045302 (2007).


\bibitem{HWWeng1} Kane, C. L. \& Mele E. J.
Z2 Topological Order and the Quantum Spin Hall Effect.
\textit{Phys. Rev. Lett.} \textbf{95}, 146802 (2005).

\bibitem{YXia} Xia, Y. \textit{et al}.
Observation of a large-gap topological-insulator class with a single Dirac cone on the surface.
\textit{Nat. Phys.} \textbf{5}, 398 (2009).

\bibitem{HWWeng1} Weng, H. M., Zhao, J. Z., Wang, Z. J., Fang, Z. \& Dai, X.
Topological crystalline Kondo insulator in mixed valence Ytterbium Borides.
\textit{Phys. Rev. Lett.} \textbf{112}, 016403 (2014).





\bibitem{ZhangXL} Zhang, X. L., Liu, L. F. \& Liu, W. M.
Quantum anomalous Hall effect and tunable topological states in 3d transition metals doped silicene.
\textit{Scientific Reports} \textbf{3}, 2908 (2013).

\bibitem{ZHQiao} Qiao, Z. H., Tse, W., Jiang, H., Yao, Y. G. \& Niu, Q.
Two-Dimensional topological insulator state and topological phase transition in bilayer graphene.
\textit{Phys. Rev. Lett.} \textbf{107}, 256801 (2011).

\bibitem{JianminZhang} Zhang, J. M., Zhu, W. G., Zhang, Y., Xiao, D. \& Yao, Y. G.
Tailoring magnetic doping in the topological insulator Bi2Se3.
\textit{Phys. Rev. Lett.} \textbf{109}, 266405 (2012).


\bibitem{silicene_Yao} Liu, C. C., Feng, W. X. \& Yao, Y. G.
Quantum spin Hall effect in silicene and two-dimensional germanium.
\textit{Phys. Rev. Lett.} \textbf{107}, 076802 (2011).




\bibitem{wanxiang_feng_MoS2} Xiao, D., Liu, G. B., Feng, W. X., Xu, X. D. \& Yao, W.
Coupled spin and valley physics in monolayers of MoS2 and other group-VI dichalcogenides.
\textit{Phys. Rev. Lett.} \textbf{108}, 196802 (2012).

\bibitem{Yang} Yang, M. \& Liu, W. M.
The d-p band-inversion topological insulator in bismuth-based skutterudites.
\textit{Scientific Reports} \textbf{4}, 5131 (2014).

\bibitem{xiangtan} Liu, W. L. \textit{et al}.
Anisotropic interactions and strain-induced topological phase transition in Sb2Se3 and Bi2Se3.
\textit{Phys. Rev. B} \textbf{84}, 245105 (2011).



\bibitem{Rliao} Liao, R., Yu, Y. X. \& Liu, W. M.
Tuning the Tricritical Point with Spin-Orbit Coupling in Polarized Fermionic Condensates.
\textit{Phys. Rev. Lett.} \textbf{108}, 080406 (2012).

\bibitem{Graphene_Yevick} Marianetti, C. A. \& Yevick, H. G.
Failure Mechanisms of Graphene under Tension.
\textit{Phys. Rev. Lett.} \textbf{105}, 245502 (2010).

\bibitem{Graphene_Chensi} Si, C., Duan, W. H., Liu, Z. \& Liu, F.
Electronic Strengthening of Graphene by Charge Doping.
\textit{Phys. Rev. Lett.} \textbf{109}, 226802 (2012).

\bibitem{Graphene_Woo} Woo, S. J. \& Son, Y. W.
Ideal strength of doped graphene.
\textit{Phys. Rev. B} \textbf{87}, 075419 (2013).

\bibitem{flexibility_phosphorene} Wei, Q. \& Peng, X. H.
Superior mechanical flexibility of phosphorene and few-layer black phosphorus.
\textit{Appl. Phys. Lett. } \textbf{104}, 251915 (2014)

\bibitem{Poisson_nanowires} McCarthy, E. K., Bellew, A. T., Sader, J. E. \& Boland, J. J.
Poisson¡¯s ratio of individual metal nanowires.
\textit{Nat. Commun.} \textbf{5}, 4336 (2014).

\bibitem{Negative_Poisson} Jiang, J. W., \& Park, H. S.
Negative poisson¡¯s ratio in single-layer black phosphorus.
\textit{Nat. Commun.} \textbf{5}, 4727 (2014).


\bibitem{arsenic_bulk} Smith, P. M., Leadbetter A. J. \& Apling, A. J.
The structures of orthorhombic and vitreous arsenic.
\textit{Philosophical Magazin} \textbf{31}, 57 (1975).


\bibitem{HSTao} Tao, H. S., Chen, Y. H., Lin, H. F., Liu, H. D. \&  Liu, W. M.
Layer Anti-Ferromagnetism on Bilayer Honeycomb Lattice.
\textit{Scientific Reports} \textbf{4}, 5367 (2014).


\bibitem{c_nanotube1} Jeroen, W. G., \textit{et al}.
Electronic structure of atomically resolved carbon nanotubes.
\textit{Nature} \textbf{391}, 59 (1997).

\bibitem{Graphene_exfoliation1} Su, C. Y., \textit{et al}.
High-Quality Thin Graphene Films from Fast Electrochemical Exfoliation.
\textit{ACS Nano} \textbf{5}, 2332 (2011).

\bibitem{Graphene_exfoliation2} Martinez, A., Fuse, K. \& Yamashita,S.
Mechanical exfoliation of graphene for the passive mode-locking of fiber lasers.
\textit{Appl. Phys. Lett.} \textbf{99}, 121107 (2011).

\bibitem{Phosphorene_exfoliation} Liu, H., \textit{et al}.
Phosphorene: An Unexplored 2D Semiconductor with a High Hole Mobility.
\textit{ACS Nano} \textbf{8}, 4033 (2014).

\bibitem{Dora} Dora, B. \& Moessner, R.
Dynamics of the spin Hall effect in topological insulators and graphene.
\textit{Phys. Rev. B} \textbf{83}, 073403 (2011).

\bibitem{QKXue1} Cheng, P. \textit{et al}.
Landau quantization of topological surface states in Bi2Se3.
\textit{Phys. Rev. Lett.} \textbf{105}, 076801 (2010).

\bibitem{phonopy} Togo, A. \textit{et al}.
First-principles calculations of the ferroelastic transition between rutile-type and CaCl2-type SiO2 at high pressures.
\textit{Phys. Rev. B} \textbf{78}, 134106 (2008).

\bibitem{QE} Giannozzi, P.  \textit{et al}.
Quantum Espresso: a modular and open-source software project for quantum simulations of materials.
\textit{J. Phys. Condens. Matter} \textbf{21}, 395502 (2009).


\bibitem{VASP2} Kresse, G.  \& Furthmuller, J.
Efficiency of ab-initio total energy calculations for metals and semiconductors using a plane-wave basis set.
\textit{Comput. Mater. Sci.} \textbf{6},  15 (1996).

\end{thebibliography}
\end{document}